\documentclass[conference]{IEEEtran}
\usepackage{cite}
\usepackage{times}
\usepackage{booktabs}
\usepackage{algorithm}
\usepackage{algorithmic}
\usepackage{enumitem}
\usepackage{multirow}
\usepackage{array}
\usepackage{amsfonts}
\usepackage{amsmath}
\usepackage{amsthm}
\usepackage{comment}
\usepackage{graphicx}
\newtheorem{theorem}{Theorem}

\begin{document}
\title{Billion-scale Network Embedding with Iterative Random Projection}




%
\author{\IEEEauthorblockN{Ziwei Zhang,
Peng Cui,
Haoyang Li,
Xiao Wang and
Wenwu Zhu}
\IEEEauthorblockA{Department of Computer Science and Technology, Tsinghua University, China\\
zw-zhang16@mails.tsinghua.edu.cn, cuip@tsinghua.edu.cn, lihaoyang96@gmail.com
\\ wangxiao007@mail.tsinghua.edu.cn, wwzhu@tsinghua.edu.cn
}
}


\maketitle

\begin{abstract}
    Network embedding, which learns low-dimensional vector representation for nodes in the network, has attracted considerable research attention recently.
    However, the existing methods are incapable of handling billion-scale networks, because they are computationally expensive and, at the same time, difficult to be accelerated by distributed computing schemes.
    To address these problems, we propose RandNE (Iterative Random Projection Network Embedding), a novel and simple billion-scale network embedding method. Specifically, we propose a Gaussian random projection approach to map the network into a low-dimensional embedding space while preserving the high-order proximities between nodes. To reduce the time complexity, we design an iterative projection procedure to avoid the explicit calculation of the high-order proximities. Theoretical analysis shows that our method is extremely efficient, and friendly to distributed computing schemes without any communication cost in the calculation. We also design a dynamic updating procedure which can efficiently incorporate the dynamic changes of the networks without error aggregation. Extensive experimental results demonstrate the efficiency and efficacy of RandNE over state-of-the-art methods in several tasks including network reconstruction, link prediction and node classification on multiple datasets with different scales, ranging from thousands to billions of nodes and edges.
\end{abstract}


%

\section{Introduction}
Network embedding is an emerging research topic in recent years, aiming to represent nodes by low-dimensional vectors while maintaining the structures and properties of the network \cite{cui2017survey}. Many methods have been proposed for network embedding, such as using random walks \cite{perozzi2014deepwalk}, matrix factorization \cite{zhang2018arbitrary} and deep learning \cite{wang2016structural}. With these methods, many network analysis tasks can be fulfilled in vector spaces and benefit from off-the-shelf machine learning models.

Despite such progress, the targeted networks of the existing methods are often in thousand or million scale. However, many real networks have billions of nodes and edges, such as social networks, e-commerce networks and the Internet. The billion-scale networks pose great computational challenges to the existing methods. The bottleneck lies in that the existing methods are all learning-based and thus involve computationally expensive optimization procedures. For example, Stochastic Gradient Descend (SGD) is a commonly used optimization method in network embedding \cite{tang2015line}, but it requires a great number of iterations to converge, which may not be feasible for billion-scale networks. One way to accelerate is to resort to distributed computing solutions, but the optimization methods, like SGD, often require global embedding information for searching gradients, leading to intense communication cost. As a result, how to design an efficient and effective billion-scale network embedding method that is friendly to distributed computing is still an open problem.

Different from learning-based methods, random projection is a simple and powerful technique to form low-dimensional embedding spaces while preserving the structures of the original space. It is also friendly to distributed computing, and thus widely exploited in large-scale data scenarios \cite{vempala2005random}. But the extremely sparse structures of real networks pose great challenges to applying random projection to network embedding. The existing work \cite{cui2017survey} has demonstrated that high-order proximities between nodes are essential to be preserved in network embedding and can effectively address the sparsity issue. Hence, how to design a high-order proximity preserved random projection method while maintaining the efficiency of the method is the key problem of billion-scale network embedding.

In this paper, we propose RandNE\footnote{The code is available at https://github.com/ZW-ZHANG/RandNE.} (Iterative Random Projection Network Embedding), a novel and simple billion-scale network embedding method based on high-order proximity preserved random projection. Specifically, we propose using Gaussian random projection to minimize the matrix factorization objective function of preserving the high-order proximity. In order to avoid the explicit calculation of high-order proximities which induces high computational complexities, we design an iterative projection procedure, enabling arbitrary high-order proximity preserved random projection with a linear time complexity. Theoretical analysis is provided to guarantee that i) RandNE is much more computationally efficient than the existing methods, ii) it can well support distributed computing without any communication cost between different servers in the calculation, and iii) it can efficiently incorporate the dynamic changes of the networks without error aggregation. These merits make RandNE a promising solution for billion-scale network embedding, even in dynamic environments.

Extensive experiments are conducted in network reconstruction, link prediction and node classification tasks on multiple datasets with different scales, ranging from thousands to billions of nodes and edges. The results show that RandNE can boost the efficiency of network embedding by about 2 orders of magnitude over state-of-the-art methods\footnote{These algorithms are tested using the source code published by their authors.} while achieving a superior or comparable accuracy. For the WeChat\footnote{One of the largest social network platforms in China.} network with 250 millions nodes and 4.8 billion edges, RandNE can produce 512-dimensional embeddings within 7 hours with 16 distributed servers.

The contributions of our paper are summarized as follows:
\begin{itemize}
\item We propose RandNE, a novel and simple random projection based network embedding method that enables billion-scale network embedding.
\item We design an iterative projection procedure to realize high-order proximities preserved random projection efficiently without explicitly calculating the high-order proximities.
\item We theoretically and empirically prove that RandNE can well support distributed computing without communication cost and can efficiently deal with dynamic networks without error aggregation.
\end{itemize}
The rest of this paper is organized as follows. In Section 2, we briefly review related works. We give our problem formulation in Section 3 and introduce our proposed method in Section 4. Experimental results are reported in Section 5. Finally, we summarize in Section 6.

\section{Related Work}
Network embedding has attracted considerable research attention in the past few years, aiming to bridge the gap between network analysis and off-the-shelf machine learning techniques. Here, we briefly review some representative network embedding methods, and readers are referred to \cite{cui2017survey} for a comprehensive survey.

The flourish of network embedding research begins when DeepWalk \cite{perozzi2014deepwalk} first proposes using truncated random walks to explore the network structure and utilizes the skip-gram model \cite{mikolov2013efficient} from word embedding to derive the embedding vectors of nodes. LINE \cite{tang2015line} takes a similar idea with an explicit objective function by setting the walk length as one, and introduces the negative sampling strategy \cite{mikolov2013distributed} to accelerate the training procedure. Node2vec \cite{grover2016node2vec} generalizes these two methods by taking potentially biased random walks for more flexibility. These random walks based methods are proven equivalent to factorizing a high-order proximity matrix \cite{qiu2018network}.

On the other hand, explicit matrix factorization methods have been proposed for network embedding. GraRep \cite{cao2015grarep} directly applies SVD to preserve high-order proximity matrices. HOPE \cite{ou2016asymmetric} proposes using generalized SVD to preserve the asymmetric transitivity in directed networks. Community structure, an important mesoscopic structure of the network, is preserved by non-negative matrix factorization in \cite{wang2017community}. \cite{chen2017fast} introduces a unified framework for matrix factorization and utilizes a sparsification technique to speed up SVD. Another approximate matrix factorization technique is introduced in \cite{yang2017fast}. AROPE \cite{zhang2018arbitrary} improves these works by preserving arbitrary-order proximity simultaneously.

Deep learning model is also applied to network embedding. SDNE \cite{wang2016structural} first considers the high non-linearity in network embedding and proposes a deep auto-encoder to preserve the first two order proximities. DHNE \cite{tu2018structural} extends this framework for preserving the indecomposability in hyper-networks.

Besides static networks, how to embed dynamic networks where nodes and edges change over time also attracts research attention. DHPE \cite{zhu2018high} and DANE \cite{li2017attributed} propose using matrix perturbation to handle the changes of edges. DepthLGP \cite{ma2018depthlgp} adopts a Gaussian process to handle out-of-sample nodes. DynamicTriad \cite{zhou2018dynamic} considers the triangle closure characteristic of network evolving.

Despite their remarkable performance, these methods are all learning-based and the targeted networks are often in thousand or million scale.
In \cite{zhou2017scalable}, a modification of DeepWalk is applied to a billion-scale network aliItemGraph. However, their method has the same time complexity as DeepWalk, which is much more computationally expensive than our method by over two orders of magnitude (see Figure \ref{fig:time} in the experiments section). Besides, it does not address the problem of distributed computing or handling dynamic networks.

How to embed networks with side information is also explored. For instance, \cite{dong2017metapath2vec,chen2017task} utilize metapaths to embed heterogenous information networks where node and edge types are available. Node attributes and node labels are taken into consideration in \cite{yang2015network,li2017attributed} and \cite{tu2016max,pan2016tri,yang2016revisiting} respectively. In this paper, we focus on the most fundamental case that only the network structure is available.

Another closely related topic is random projection \cite{vempala2005random,arriaga2006algorithmic,shi2012margin,choromanski2017unreasonable}, which is widely adopted in dimension reduction. But existing random projection methods do not consider the sparsity problem in network embedding, and thus cannot be directly applied. 

\section{Notations and Problem Formulation}
\subsection{Notations}
First, we summarize the notations used in this paper. For a network $G$ with $N$ nodes and $M$ edges, we use $\mathbf{A}$ to denote the adjacency matrix. In this paper, we mainly consider undirected networks, so $\mathbf{A}$ is symmetric. $\mathbf{A}(i,:)$ and $\mathbf{A}(:,j)$ denote its $i^{th}$ row and $j^{th}$ column respectively. $\mathbf{A}(i,j)$ is the element in the $i^{th}$ row and $j^{th}$ column. $\mathbf{A}^T$ denotes the transpose of $\mathbf{A}$. Throughout the paper, we use bold uppercase characters to denote matrices and bold lowercase characters to denote vectors, e.g. $\mathbf{X}$ and $\mathbf{x}$ respectively. We use dot to denote the matrix product of two matrices, e.g. $\mathbf{B} \cdot \mathbf{C}$. Functions are marked by curlicue, e.g. $\mathcal{F}(\cdot)$.

\subsection{Problem Formulation}
To represent nodes in a network by low-dimensional vectors, one commonly adopted objective function in network embedding is matrix factorization, which decomposes a targeted similarity function of the adjacency matrix $\mathcal{F}(\mathbf{A}) \in \mathbb{R}^{N \times N}$ into the product of two low-dimensional matrices $\mathbf{U},\mathbf{V} \in \mathbb{R}^{N \times d}$ with the following objective function:
\begin{equation}\label{eq:obj1}
    \min_{\mathbf{U},\mathbf{V}} \left\| \mathcal{F}(\mathbf{A})- \mathbf{U} \cdot \mathbf{V}^T \right\|_p,
\end{equation}
where $p$ is the norm and $d$ is the dimensionality of the embedding. In this paper, we only consider undirected networks and symmetric similarities, so $\mathbf{U} = \mathbf{V}$. We also focus on the spectral norm, i.e. $p=2$, which is widely adopted \cite{liberty2013simple}. The adjacency matrix $\mathbf{A}$ could be replaced by other variants, such as the Laplacian matrix or transition matrix \cite{qiu2018network}. Here, we focus on the adjacency matrix unless stated otherwise.

The previous work has shown that high-order proximities are essential to be preserved in network embedding, which can be formulated as a polynomial function of the adjacency matrix \cite{yang2017fast,chen2017fast}. In this paper, we assume that $\mathcal{F}(\mathbf{A})$ is a positive semi-definite function, so it can be formulated as $\mathcal{F}(\mathbf{A}) = \mathbf{S} \cdot \mathbf{S}^T$. Then, we can rewrite Eq. \eqref{eq:obj1} as:
\begin{equation}\label{eq:obj2}
\begin{gathered}
    \min_{\mathbf{U}} \left\| \mathbf{S} \cdot \mathbf{S}^T - \mathbf{U} \cdot \mathbf{U}^T \right\|_2\\
     \mathbf{S} = \alpha_0 \mathbf{I} + \alpha_1 \mathbf{A} + \alpha_2 \mathbf{A}^2 + ... + \alpha_q \mathbf{A}^q,
\end{gathered}
\end{equation}
where $\mathbf{S}$ is the high-order proximity matrix, $\alpha_0, \alpha_1, ... ,\alpha_q$ are pre-defined weights and $q$ is the order.

From Eckart-Young theorem \cite{eckart1936approximation}, it is well known that Singular Value Decomposition (SVD) can lead to the optimal solution of Eq. \eqref{eq:obj2}. However, SVD is computationally expensive and thus not suitable for large-scale networks.

\section{RandNE: the Proposed Method}
\subsection{Gaussian Random Projection Embedding}
To minimize the objective function in Eq. \eqref{eq:obj2}, an extremely simple yet effective method is random projection, and Gaussian random projection is widely used \cite{vempala2005random}. Formally, let $\mathbf{R} \in \mathbb{R}^{N \times d}$ and each element of $\mathbf{R}$ follows an i.i.d Gaussian distribution $\mathbf{R}(i,j) \sim \mathcal{N}\left(0,\frac{1}{d} \right)$. Then, the embeddings $\mathbf{U}$ can be obtained by performing a matrix product:
\begin{equation}\label{eq:emb}
    \mathbf{U} = \mathbf{S} \cdot \mathbf{R} = \left( \alpha_0 \mathbf{I} + \alpha_1 \mathbf{A} + \alpha_2 \mathbf{A}^2 + ... + \alpha_q \mathbf{A}^q \right) \mathbf{R},
\end{equation}
i.e. we randomly project the proximity matrix $\mathbf{S}$ into a low-dimensional subspace. Gaussian random projection has an theoretical guarantee, as we specific in the following theorem.
\begin{theorem}\label{theorem:bound}
For any similarity matrix $\mathbf{S}$, denote its rank as $r_{\mathbf{S}}$. Then, for any $\epsilon \in \left(0,\frac{1}{2} \right)$, the following equation holds:
\begin{equation}
    P \left[ \; \left\| \mathbf{S} \cdot \mathbf{S}^T - \mathbf{U} \cdot \mathbf{U}^T \right\|_2 > \epsilon \left\| \mathbf{S}^T \cdot \mathbf{S} \right\|_2 \right] \leq 2 r_{\mathbf{S}} e^{-\frac{\left( \epsilon^2 - \epsilon^3 \right) d}{4}  },
\end{equation}
where $\mathbf{U} = \mathbf{S} \cdot \mathbf{R}$ and $\mathbf{R}$ is a Gaussian random matrix.
\end{theorem}

\begin{IEEEproof}
See appendix.
\end{IEEEproof}

The theorem basically shows that the residual of our projection $\mathbf{S} \cdot \mathbf{S}^T - \mathbf{U} \cdot \mathbf{U}^T$ has a much smaller spectral radius compared to the spectral radius of the original high-order proximity $\mathbf{S} \cdot \mathbf{S}^T$. In other words, the embedding $\mathbf{U}$ captures the ``core component'' of the high-order proximity. As a result, performing a Gaussian random projection can effectively minimize the objective function in Eq. \eqref{eq:obj2}. Actually, Gaussian random projection is also known to have other merits, such as preserving the margin for classification \cite{shi2012margin}, which we omit for brevity.

For the projection matrix, it is proven that orthogonal Gaussian random matrix can further improve the accuracy, which can be easily obtained by performing a Gram Schmidt process on the columns of the Gaussian random matrix \cite{choromanski2017unreasonable}. In this paper, we use the orthogonal Gaussian random matrix as the projection matrix.

However, since $\mathbf{S}$ may not be a sparse matrix, directly calculating $\mathbf{S}$ and performing the projection is still time consuming and not scalable to large-scale networks.

\subsection{Iterative Projection}
To address the efficiency problem, we design an iterative projection procedure to avoid the explicit calculation of the high-order proximity matrix $\mathbf{S}$. Specifically, from Eq. \eqref{eq:emb}, we can decompose $\mathbf{U}$ into matrices of different orders:
\begin{equation}\label{eq:combine}
    \mathbf{U} =  \alpha_0 \mathbf{U}_0 + \alpha_1 \mathbf{U}_1 + ... + \alpha_q \mathbf{U}_q,
\end{equation}
where $\mathbf{U}_i = \mathbf{A}^i \cdot \mathbf{R}, 0\leq i \leq q$. Then, the decomposed parts, $\mathbf{U}_1 ...\mathbf{U}_q$, can be calculated iteratively:
\begin{equation}\label{eq:irp}
     \mathbf{U}_i = \mathbf{A} \cdot \mathbf{U}_{i-1} \ , \ \forall 1 \leq i \leq q.
\end{equation}Note that in Eq. \eqref{eq:irp}, we only need to calculate the matrix product of the adjacency matrix and a low-dimensional matrix. Since the adjacency matrix is sparse, we can use sparse matrix products, which are highly scalable and efficient.

    \begin{algorithm}[t]
    \caption{RandNE: Iterative Random Projection Network Embedding}\label{alg1}
    \begin{algorithmic}[1]
    \REQUIRE Adjacency Matrix $\mathbf{A}$, Dimensionality $d$, Order $q$, Weights $\alpha_0,\alpha_1,...,\alpha_q$
    \ENSURE Embedding Results $\mathbf{U}$
    \STATE Generate $\mathbf{R} \in \mathbb{R}^{N \times d} \sim \mathcal{N}(0,\frac{1}{d})$ \label{step1}
    \STATE Perform a Gram Schmidt process on $\mathbf{R}$ to obtain the orthogonal projection matrix $\mathbf{U}_0$ \label{step2}
    \FOR{i in 1:q}\label{step3}
        \STATE Calculate $\mathbf{U}_i = \mathbf{A} \cdot \mathbf{U}_{i-1}$
    \ENDFOR\label{step4}
    \STATE Calculate $\mathbf{U} = \alpha_0 \mathbf{U}_0 + \alpha_1 \mathbf{U}_1 + ... + \alpha_q \mathbf{U}_q$ \label{step5}
    \end{algorithmic}
    \end{algorithm}

\subsection{Time Complexity and Distributed Computing}
We show our algorithm framework in Algorithm \ref{alg1}. Then, we analyze the time complexity of Algorithm \ref{alg1}. The complexity of line \ref{step1} is $O(N\cdot d)$, the complexity of line \ref{step2} is $O(N\cdot d^2)$, the complexity of each iteration from line \ref{step3} to line \ref{step4} is $O(M \cdot d)$ and the complexity of line \ref{step5} is $O(q \cdot N \cdot d)$, where $N$ and $M$ are the number of nodes and edges in the network respectively, $q$ is the preset order and $d$ is the dimensionality of the embedding space. As a result, the overall time complexity is $O\left(N \cdot d^2 + M \cdot q \cdot d\right)$, i.e. our method is linear with respect to the network size.

From the above analysis, we can also see that our method is extremely efficient because it only needs to iterate $q$ times, and within each iteration, only a simple matrix product needs to be calculated. In contrast, although some existing network embedding methods are also proven to have linear time complexities, such as the embedding methods based on SGD \cite{tang2015line} or SVD \cite{ou2016asymmetric}, they inevitably need dozens or hundreds of iterations in the optimization. As a result, our method is more efficient than these methods by orders of magnitude, and is thus more suitable for billion-scale network embedding.

In addition, according to the property of matrix products, each dimension (i.e. column) of $\mathbf{U}_i$ can be calculated separately without any information from other dimensions. We formalize this property in the following theorem.
\begin{theorem}\label{The:distributed}
    For any $j \neq l$, the calculation of $\mathbf{U}_i(:,j)$ and $\mathbf{U}_i(:,l), 1 \leq i \leq q$ from line \ref{step3} to line \ref{step4} in Algorithm \ref{alg1} are independent, if $\mathbf{A}$ and $\mathbf{U}_0$ are known to all the servers.
\end{theorem}
\begin{IEEEproof}
Straightforward from the property of matrix products.
\end{IEEEproof}
The theorem shows that our method naturally supports distributed computing by allocating the calculation of different dimensions into distributed servers, and no communication is needed during the calculation process, if $\mathbf{A}$ and $\mathbf{U}_0$ are known to all the servers.
We design a simple distributed protocol using the theorem, as specified in Algorithm \ref{alg3}.
    \begin{algorithm}[t]
    \caption{Distributed Calculation of RandNE}\label{alg3}
    \begin{algorithmic}[1]
    \REQUIRE Adjacency matrix $\mathbf{A}$, Initial Projection $\mathbf{U}_0$, Parameters of RandNE, $K$ Distributed Servers
    \ENSURE Embedding Results $\mathbf{U}$
    \STATE Broadcast $\mathbf{A}$, $\mathbf{U}_0$ and parameters into $K$ servers
    \STATE Set i = 1
    \REPEAT \IF{There is an idle server $k$}
              \STATE Calculate $\mathbf{U}(:,i)$ in server $k$
              \STATE Set i = i + 1
              \STATE Gather $\mathbf{U}(:,i)$ from server $k$ after calculation
              \ENDIF
    \UNTIL{$i > d$}
    \STATE Return $\mathbf{U}$
    \end{algorithmic}
    \end{algorithm}

For networks that cannot be stored in the memory or when the number of servers exceeds the dimensionality, we can use more advanced distributed matrix multiplication algorithms, such as \cite{vastenhouw2005two,boman2013scalable}, which we leave as the future work.
This is in contrast with the existing methods which can only be parallelizable within one server but are hard to be distributed because of communication cost. This merit lays another foundation for applying RandNE to billion-scale networks.

\subsection{Dynamic Updating}\label{sec:update}
As many real networks are dynamic, we next show how to efficiently update RandNE to incorporate dynamic changes.

First, we focus on the changes of edges. From Algorithm \ref{alg1}, to update the final embedding vectors, we only need to update the decomposed parts $\mathbf{U}_i, 1 \leq i \leq q$. Formally, we denote the changes in the adjacency matrix as $\Delta \mathbf{A}$ and the changes in $\mathbf{U}_i$ as $\Delta \mathbf{U}_i, 1\leq i \leq q$. From Eq. \eqref{eq:irp}, we have:
\begin{equation}\label{eq:update}
\begin{gathered}
    \mathbf{U}_i + \Delta \mathbf{U}_i = \left( \mathbf{A} + \Delta \mathbf{A} \right) \cdot \left( \mathbf{U}_{i-1} + \Delta \mathbf{U}_{i-1} \right) \\
   \Rightarrow  \Delta \mathbf{U}_i = \mathbf{A} \cdot \Delta \mathbf{U}_{i-1} + \Delta \mathbf{A} \cdot \mathbf{U}_{i-1} + \Delta \mathbf{A} \cdot \Delta \mathbf{U}_{i-1}.
\end{gathered}
\end{equation}
Then, we can iteratively calculate $\Delta \mathbf{U}_i$ using Eq. \eqref{eq:update}.

Besides, nodes in the network may also be added or deleted. As for deleting nodes, it can be treated equivalently as the changes of edges by deleting all edges of the deleted nodes. For newly added nodes, we first add some empty nodes (i.e. without any edge) to make the dimensionality of the matrices match. Specifically, we denote $N'$ as the number of added nodes. For the projection matrix $\mathbf{U}_0$, we can generate an additional orthogonal Gaussian random matrix $\hat{\mathbf{U}}_0 \in \mathbb{R}^{N' \times d}$ and concatenate it with the current projection matrix $\mathbf{U}_0$ to form the new projection matrix $\mathbf{U}_0'$. For other $\mathbf{U}_i,1\leq i \leq q$, we can easily find that they have all-zero elements for the empty nodes, i.e. we only need to add $N'$ all-zero rows to match the dimensionality. Then, we can add edges to those newly added nodes using Eq. \eqref{eq:update}.

  \begin{algorithm}[t]
    \caption{Dynamic Updating of RandNE}\label{alg2}
    \begin{algorithmic}[1]
    \REQUIRE Adjacency Matrix $\mathbf{A}$, Dynamic Changes $\Delta \mathbf{A}$, Previous Projection Results $\mathbf{U}_0,\mathbf{U}_1,...,\mathbf{U}_q$
    \ENSURE Updated Projection Results $\mathbf{U}^\prime_0,\mathbf{U}^\prime_1,...,\mathbf{U}^\prime_q$
    \IF{$\Delta \mathbf{A}$ includes $N'$ new nodes} \label{step11}
        \STATE Generate an orthogonal projection $\hat{\mathbf{U}}_0 \in \mathbb{R}^{N' \times d}$
        \STATE Concatenate $\hat{\mathbf{U}}_0$ with $\mathbf{U}_0$ to obtain $\mathbf{U}^\prime_0$
        \STATE Add $N'$ all-zero rows in $\mathbf{U}_1...\mathbf{U}_q$
    \ENDIF \label{step12}
    \STATE Set $\Delta \mathbf{U}_0 = 0$ \label{step13}
    \FOR{i in 1:q}
        \STATE Calculate $\Delta \mathbf{U}_i$ using Eq. \eqref{eq:update}
        \STATE Calculate $\mathbf{U}^\prime_i = \mathbf{U}_i + \Delta \mathbf{U}_i$
    \ENDFOR \label{step14}
    \end{algorithmic}
    \end{algorithm}

We show the framework of dynamic updating in Algorithm \ref{alg2}. As only local changes are involved, the updating is computationally efficient, as we specific in the following theorem.
\begin{theorem}\label{thm3}
The time complexity of Algorithm \ref{alg2} is linear with respect to the number of changed nodes and number of changed edges respectively.
\end{theorem}
\begin{IEEEproof}
See appendix.
\end{IEEEproof}

Another merit of our updating method is that it has no error aggregation, i.e. the dynamic updating algorithm leads to the same results as rerunning the static algorithm. We formalize this property in the following theorem.
\begin{theorem}\label{thm4}
Denote $\mathbf{A}_0$ and $\Delta \mathbf{A}_1, \Delta \mathbf{A}_2,..., \Delta \mathbf{A}_t$ as the initial adjacency matrix and its dynamic changes in $t$ time steps respectively. Denote $\mathbf{U}$ as the final embedding results of applying Algorithm \ref{alg1} to $\mathbf{A}_0$ and then updating $t$ times using $\Delta \mathbf{A}_1, \Delta \mathbf{A}_2,..., \Delta \mathbf{A}_t$ and Algorithm \ref{alg2}. Denote $\mathbf{U}'$ as the embedding results of applying Algorithm \ref{alg1} to $\mathbf{A} = \mathbf{A}_0 + \Delta \mathbf{A}_1 + \Delta \mathbf{A}_2 + ... + \Delta \mathbf{A}_t$. If $\mathbf{U}$ and $\mathbf{U}'$ are calculated using the same hyper-parameters and random seed, then $\mathbf{U} = \mathbf{U}'$.
\end{theorem}
\begin{IEEEproof}
Since the same random seed is used, $\mathbf{U}_0$ = $\mathbf{U}_0'$. Then, applying Eqs. \eqref{eq:combine} \eqref{eq:irp} \eqref{eq:update} leads to the results.
\end{IEEEproof}

Combining Theorem \ref{thm3} and Theorem \ref{thm4}, our updating method can effectively incorporate the dynamic changes of networks with high computational efficiency.

\section{Experiments}
\subsection{Experimental Setting}
To comprehensively evaluate the efficacy of RandNE, we first conduct experiments on 3 moderate-scale social networks\footnote{http://socialcomputing.asu.edu/pages/datasets}: BlogCatalog, Flickr, Youtube, and then evaluate our method on a billion-scale network, WeChat. The statistics of the datasets are summarized in Table \ref{Datasets}.
    \begin{table}
    \centering
    \caption{The Statistics of Datasets}\label{Datasets}
    \begin{tabular}{ l | c | c | c }
    \hline
    Dataset & \# Nodes & \# Edges & \# Labels \\ \hline
    BlogCatalog & 10,312 & 667,966   & 39 \\ \hline
    Flickr      & 80,513 & 11,799,764 &  47 \\ \hline
    Youtube     & 1,138,499 & 5,980,886 &  195 \\ \hline
    WeChat        & 250 million & 4.8 billion & - \\ \hline
    \end{tabular}
    \end{table}


We compare our method with the following baselines:
\begin{itemize}
\item DeepWalk \cite{perozzi2014deepwalk}\footnote{https://github.com/phanein/deepwalk} uses random walks and the skip-gram model to learn embeddings. We use two parameter settings: one suggested in the paper and one used in the implementation of the authors, and report the best results.
\item LINE \cite{tang2015line}\footnote{https://github.com/tangjianpku/LINE} explicitly preserves the first two order proximities, denoted as LINE$_{1st}$ and LINE$_{2nd}$ respectively. We exclude the results of concatenating them because no obvious improvement is observed. We use the default parameter settings except the number of training samples, which we conduct a line search for the optimal value.
\item Node2vec \cite{grover2016node2vec}\footnote{https://github.com/snap-stanford/snap} generalizes DeepWalk and LINE by using potentially biased random walks.
    We use the default settings for all parameters except the bias parameters $p,q$, which we conduct a grid search from $\left\{0.5,1,2\right\}$.
\item SDNE \cite{wang2016structural}\footnote{https://github.com/suanrong/SDNE} proposes a deep auto-encoder to preserve the first and the second order proximities simultaneously. We use the default parameter settings and the auto-encoder structure in the implementation of the authors.

\end{itemize}
There are also other methods like GraRep \cite{cao2015grarep} and M-NMF \cite{wang2017community}, but we exclude them here for their scalability issues. We also exclude the results of SDNE on Youtube because it fails to terminate in one week. On the WeChat network, as all these baselines cannot terminate within acceptable time, we mainly compare our method with other simpler graph-based methods.

For our method RandNE, we set the order $q=3$ with a grid search for the weights. Please note that the weights only affect the last step of our algorithm (i.e., line \ref{step5} in Algorithm \ref{alg1}), so our proposed method is very efficient in tuning them. For the node classification task, we use transition matrix to replace the adjacency matrix because substantial improvement is observed. All hyper-parameters of our method and the baselines are tuned using a small validation set, which we set as 10\% for moderate-scale networks and 1\% for the billion-scale network.

For all the methods, we uniformly set the dimensionality as $d = 128$ unless stated otherwise. All experiments are conducted in a single PC with 2 I7 processors and 48GB memory, except for Section \ref{sec:wechat}, where we run our method in a distributed cluster.

\subsection{Moderate-scale Networks}

\subsubsection{Running Time Comparison}
\begin{figure}
\centering
\hspace{0.6cm}
\includegraphics[width=7.5cm]{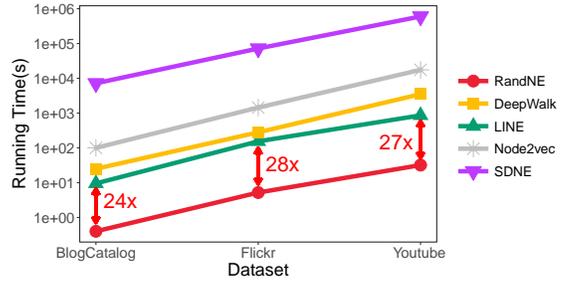}
\caption{The running time comparison of different methods. Our method RandNE can boost the efficiency by more than 24 times over state-of-the-art methods on all networks.}
\label{fig:time}
\end{figure}

To compare the efficiency of different methods, we first report the running time of all the methods in Figure \ref{fig:time}. The results show that RandNE can boost the efficiency by more than 24 times over the baselines on all networks, which is consistent to our time complexity analysis.
Note that the baselines are tested using the source code published by their authors.
We realize that there might be slight differences in the programming languages and implementation details, but the impact of these factors can be safely ignored considering the improvement of 24 times. So we directly report their results for reproducibility concerns.
The extreme efficiency lays the foundation for applying RandNE to billion-scale networks.

\subsubsection{Network Reconstruction}\label{sec:recon}
    \begin{table}
    \caption{AUC scores of Network Reconstruction}\label{table:recon}
    \centering
    \begin{tabular}{ c | c | c | c }
    \hline
    Dataset      & BlogCatalog & Flickr & Youtube \\ \hline
    RandNE       &  $\mathbf{0.958}$ & $\mathbf{0.953}$ & 0.982 \\ \hline
    DeepWalk     &  0.843 & 0.951  & 0.995 \\ \hline
    LINE$_{1st}$ &  0.901 & 0.947  & $\mathbf{0.999}$ \\ \hline
    LINE$_{2nd}$ &  0.761 & 0.936  & 0.970 \\ \hline
    Node2vec     &  0.805 & 0.890  & 0.952 \\ \hline
    SDNE         &  0.950 & 0.919  & - \\ \hline
    \end{tabular}
    \end{table}
\begin{figure*}[t]
\centering
\hspace{-0.5cm}
\includegraphics[width=12cm]{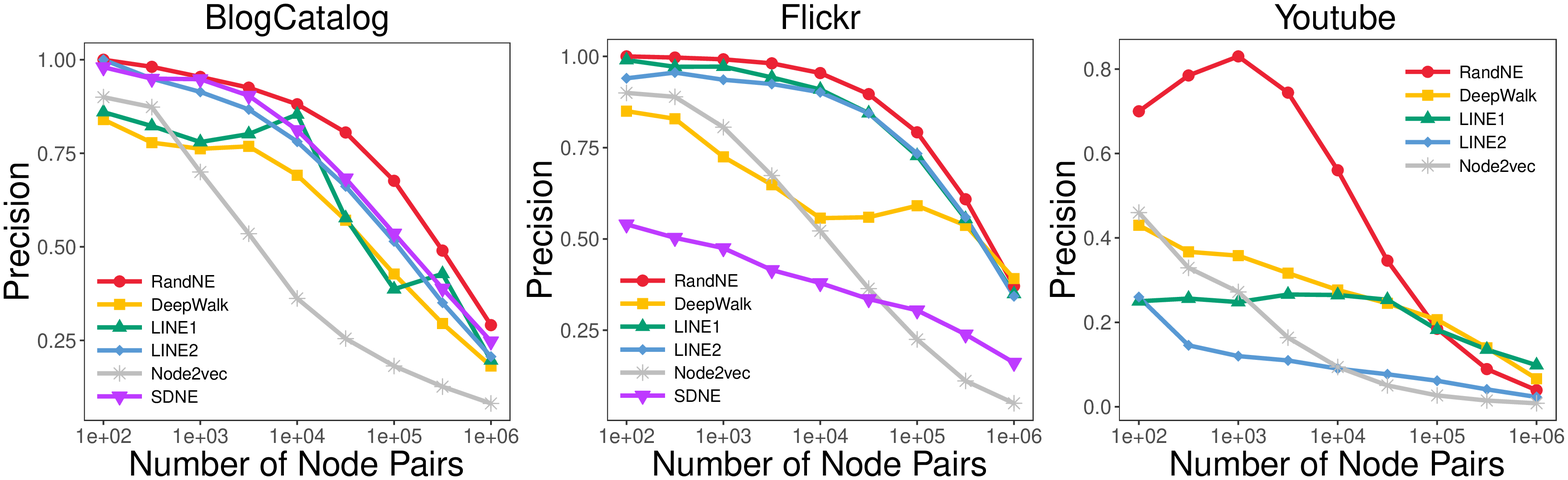}
\caption{The Precision@K of network reconstruction on moderate-scale networks. We train embedding vectors and rank pairs of nodes according to their inner-product similarities. The results show that our proposed method can well preserve the network structure and reconstruct the given network.}
\label{fig:recon}
\end{figure*}

 One basic objective for network embedding is to reconstruct the network. Specifically, we train embedding vectors and rank pairs of nodes according to their inner product similarities.
 Then, the top ranking pairs are used to reconstruct the network because large similarities indicate high probabilities of having edges.
 For the evaluation metrics, we use Area Under Curve (AUC) \cite{fawcett2006introduction} and Precision@K \cite{wang2016structural} defined as:
 \begin{equation}\label{eq:prec}
    Precision@K = \frac{1}{K} \sum_{i=1}^{K} \delta_i,
 \end{equation}
where $\delta_i = 1$ means the $i^{th}$ reconstructed pair is correct, $\delta_i = 0$ represents a wrong reconstruction and $K$ is the number of evaluated pairs. On Youtube, the number of possible pairs of nodes $\frac{N(N-1)}{2}$ is too large to evaluate, so we sample 1\% for evaluation, as in \cite{ou2016asymmetric}.

The results are shown in Table \ref{table:recon} and Figure \ref{fig:recon}. Our proposed method consistently outperforms the baselines on the metric Precision@K. On AUC, our method achieves the best performance on BlogCatalog and Flickr, and has comparable performance on Youtube. Considering the significant improvement in efficiency and the simplicity of our model, we regard the performance of RandNE in accuracy aspect to be satisfactory and somewhat beyond expectation, which well demonstrates the effectiveness of random projection in preserving network structure.

\subsubsection{Link Prediction}\label{sec:lp}
Link prediction, aiming to predict future links using the current network structure, is an important task of network embedding.
In our experiments, we randomly hide 30\% of the edges for testing. After training embedding vectors on the rest of the network, we rank pairs of nodes in a similar way as network reconstruction and evaluate the results on the testing network. The process is repeated 5 times and the average results are reported.

\begin{figure*}[t]
\centering
\hspace{-0.3cm}
\includegraphics[width=12cm]{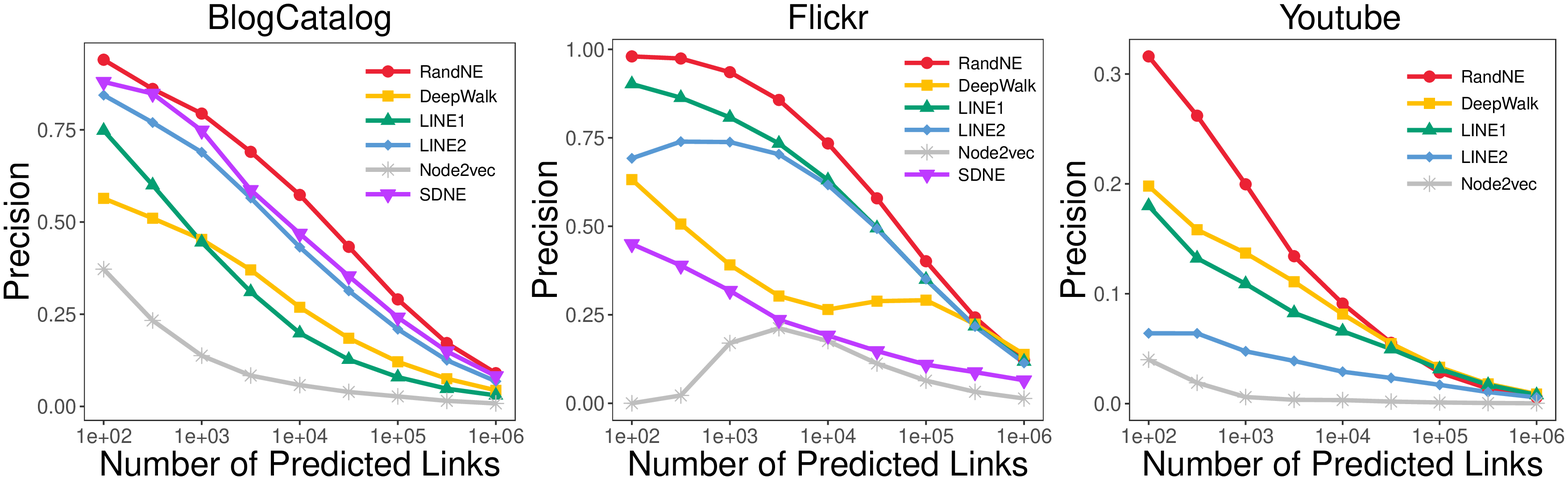}
\caption{The Precision@K of link prediction on moderate-scale networks. We randomly split the network into training and testing. After training embedding vectors on the training network, we predict links by ranking similarities of node pairs and make evaluation on the testing network. The results show that our proposed method outperforms the baselines in link prediction.}
\label{fig:lp}
\end{figure*}

From Table \ref{table:lp} and Figure \ref{fig:lp}, we can see that our proposed method still outperforms the baselines in nearly all cases except the AUC score on Youtube, as in network reconstruction. The results demonstrate that besides reconstructing the network, RandNE also has good inference abilities, which we attribute to effectively preserving the high-order proximity.

\begin{table}
    \caption{AUC scores of Link Prediction.}\label{table:lp}
    \centering
    \begin{tabular}{ c | c | c | c }
    \hline
    Dataset      & BlogCatalog & Flickr & Youtube \\ \hline
    RandNE       & $\mathbf{0.944}$  & $\mathbf{0.940}$ & 0.887 \\ \hline
    DeepWalk     & 0.760  & 0.938  &  0.909 \\ \hline
    LINE$_{1st}$ & 0.667  & 0.909  &  0.847 \\ \hline
    LINE$_{2nd}$ & 0.762  & 0.932  &  $\mathbf{0.959}$ \\ \hline
    Node2vec     & 0.650  & 0.865  &  0.778 \\ \hline
    SDNE         & 0.940  & 0.926  &  - \\ \hline
    \end{tabular}
\end{table}

\subsubsection{Node Classification}\label{sec:nodeclassification}
Node classification is a typical application of network embedding. Specifically, we follow the experimental setting in baselines and randomly split the nodes into training and testing set. Then, an one-vs-all logistic regression with L2 regularization \cite{fan2008liblinear} is trained using the embeddings on the training set, and tested on the testing set. Following \cite{tang2015line}, we normalize each row of the embedding vectors. We use two measurements, Macro-F1 and Micro-F1 \cite{perozzi2014deepwalk}, to evaluate the performance. The average results of 5 runs are reported.
\begin{figure*}[t]
\centering
\hspace{0.5cm}
\includegraphics[width=13.5cm]{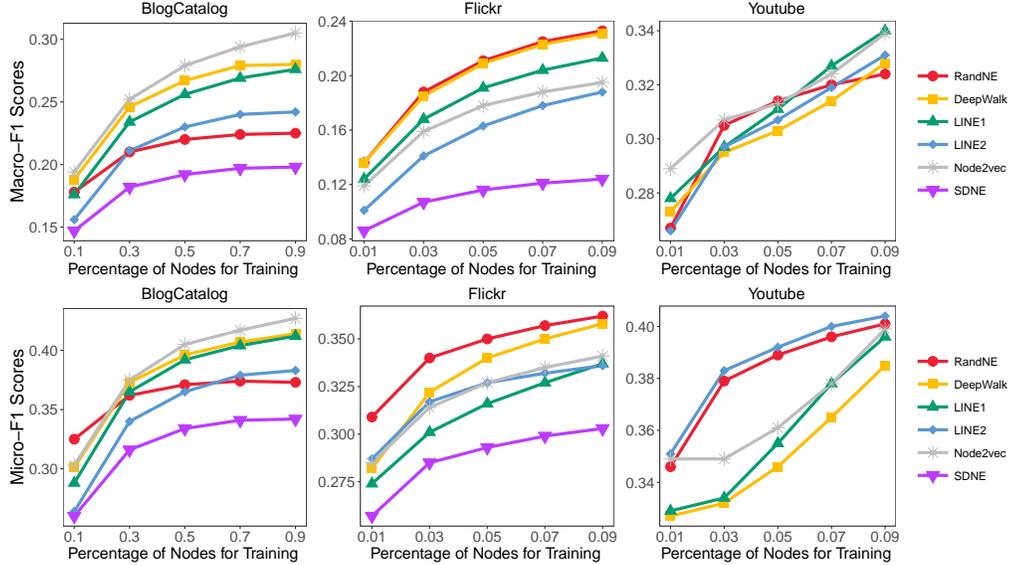}
\caption{The results of node classification on moderate-scale networks. We train an one-vs-all logistic regression on the embedding vectors as the classifier. The results show that our proposed method achieves comparable performance with the baselines.}
\label{fig:classification}
\end{figure*}

From Figure \ref{fig:classification}, different networks show different patterns in terms of node classification performance. On Flickr, RandNE achieves the best results while Node2vec and LINE show good performance on BlogCatalog and Youtube respectively. One plausible explanation for such inconsistency is that different networks have different inherent structures corresponding to the specific classification task, and no single method can dominate others on all datasets. But in general, we can safely conclude that RandNE has comparable results with the baselines in node classification while being significantly faster.

\subsubsection{Structural Role Classification}\label{sec:class}
Recently, how to preserve the structural role of nodes in network embedding attracts some research attention \cite{ribeiro2017struc2vec}, which has important applications such as influence maximization and measuring node centrality. To validate the effectiveness of our method in structural role classification, we conduct experiments on three air-traffic networks\footnote{https://github.com/leoribeiro/struc2vec} from Brazilian, European and American as in \cite{ribeiro2017struc2vec}, where the networks have 131 nodes and 2,006 edges, 399 nodes and 11,986 edges, 1,190 nodes and 27,198 edges respectively. The networks are constructed by assigning airports as nodes and airlines as edges. Each node is assigned a ground-truth label ranging from 1 to 4 to indicate the level of activities of the airports.

 The experimental setting is similar to node classification in Section \ref{sec:nodeclassification} except that we use accuracy, i.e. the percentage of nodes whose labels are correctly predicted, as the measurement since the labels have the same size. We uniformly set the dimensionality of embedding as 16 since the networks have small sizes. The average results of 20 runs are reported.

\begin{figure*}[t]
\centering
\hspace{0cm}
\includegraphics[width=13.5cm]{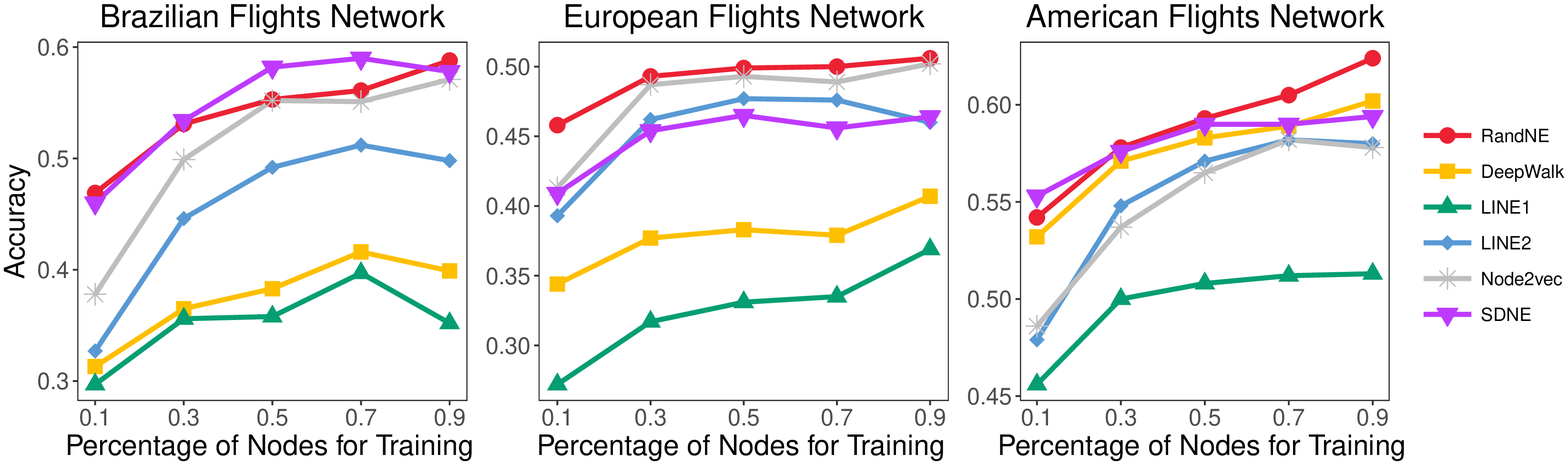}
\caption{The accuracy of structural role classification. We use embedding vectors to predict the structural role of nodes in air-traffic networks.
The results show that RandNE effectively captures the structural role of nodes.}
\label{fig:structural}
\end{figure*}

From Figure \ref{fig:structural}, we can see that RandNE consistently achieves the best results on European Flights Network. On Brazilian and American Flights Network, RandNE is only second to SDNE with tiny differences. However, RandNE is much more efficient than SDNE by about 4 orders of magnitude (see Figure \ref{fig:time}). The results demonstrate that RandNE can effectively capture the structural role of nodes.

\subsubsection{Analysis}
\begin{figure}[t]
\centering
\includegraphics[width=8.5cm]{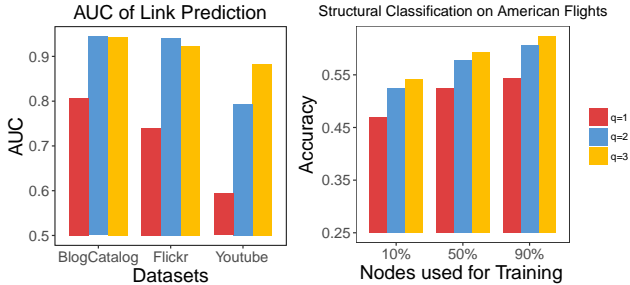}
\caption{Parameter analysis. Figure shows that the high-order proximity ($q > 1$) greatly outperforms the simple random projection ($q=1$), demonstrating the importance of preserving high-order proximities.}
\label{fig:ParaOrder}
\end{figure}

In RandNE, we use iterative random projection to preserve high-order proximities. Here we analyze the effect of the proximity order, or equivalently, the number of iterations $q$. We report the results of varying $q$ from 1 to 3 with the same experimental settings. For brevity, we only report AUC scores of link prediction and the accuracy of structural role classification on American Flights in Figure \ref{fig:ParaOrder}, while other datasets and tasks show similar patterns. The results show that iterative random projection ($q > 1$) greatly and consistently outperforms the simple random projection ($q=1$), demonstrating the importance of preserving high-order proximities in network embedding.

\begin{figure}[t]
\centering
\includegraphics[width=7.5cm]{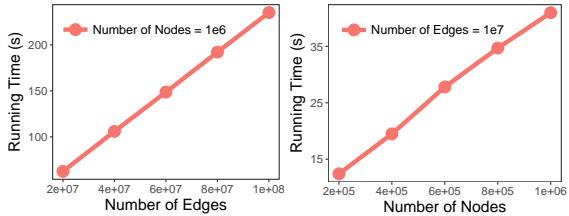}
\caption{Scalability analysis. Figure shows the linear time complexity of our method with respect to the number of nodes and number of edges respectively. }
\label{fig:ParaScale}
\end{figure}

To verify the scalability of RandNE, we conduct experiments on random networks (i.e. the Erdos Renyi model \cite{erdos1960evolution}). We record the running time when fixing the number of nodes (as one million) or fixing the number of edges (as ten million) while varying the other. Figure \ref{fig:ParaScale} shows that the running time grows linearly with respect to the number of nodes and number of edges respectively, verifying the scalability of RandNE.

Since our method is based on random projection, we also empirically analyze the impact of randomization. Specifically, we set different random seeds for RandNE with the same experimental setting, and report the mean value and standard deviation of 10 runs. For brevity, we only report the AUC scores of network reconstruction and link prediction in Table \ref{table:randomization}, while other tasks show similar patterns. The results show that our method is quite stable with respect to randomization.

\begin{table}[t]
    \caption{Mean value and standard deviation of AUC with 10 different initializations.}\label{table:randomization} 
    \centering
    \begin{tabular}{ c | c | c | c }
    \hline
    Dataset          &  BlogCatalog & Flickr & Youtube \\ \hline
    Reconstruction   &  $0.958\pm0.001$ & $0.954\pm0.003$ & $0.981\pm$ 0.002\\ \hline
    Link Prediction  &  $0.945\pm0.001$ & $0.941\pm0.002$ & $0.887\pm0.001$ \\ \hline
    \end{tabular}
    \end{table}

\subsection{A Billion-scale Network}\label{sec:wechat}
WeChat\footnote{http://www.wechat.com/en/} is one of the largest online social networks in China with more than one billion active users. We use the friendships data provided by WeChat from January 21, 2011 (the launch day of WeChat) to January 20, 2013, which contains 250 million nodes and 4.8 billion edges in total. The data is strictly anonymized for privacy purposes. Since no node label information is available, we mainly conduct experiments on network reconstruction and link prediction. As none of the aforementioned network embedding method can be applied to network of such a scale, we compare our method with two widely used graph-based measures: Common Neighbors and Adamic Adar \cite{liben2007link}. The benchmark accuracy of random guessing is also added. The experiments are conducted in a distributed cluster with 16 computing servers, where each server has 2 Xeon E5 CPU and 128GB memory. For RandNE, we set the embedding dimensionality as $d=512$.
\begin{table}[t]
    \caption{AUC scores of network reconstruction on WeChat network.}\label{table:auclarge}
    \centering
    \begin{tabular}{ c | c }
    \hline
    Method          & AUC  \\ \hline
    RandNE             & $\mathbf{0.989}$ \\ \hline
    Common Neighbors & 0.783 \\ \hline
    Adamic Adar     & 0.783 \\ \hline
    Random          & 0.500 \\ \hline
    \end{tabular}
    \end{table}
\subsubsection{Network Reconstruction}
The experimental setting is similar to moderate-scale networks (Section \ref{sec:recon}), i.e. we rank pairs of nodes according to their inner product similarities and reconstruct the network. We report the AUC scores in Table \ref{table:auclarge}. Here we omit the other metric, Precision@K, because the number of possible node pairs $\frac{N(N-1)}{2} \approx 10^{16}$ is so large that even sampling is infeasible.

The results show that our proposed method greatly outperforms Common Neighbor and Adamic Adar. A plausible reason is that our method preserves the high-order proximity information in the embedding vectors by performing the iterative random projection, while the baselines only count local proximities. Adamic Adar, as a frequency-weighted modification of Common Neighbors, has the same accuracy as Common Neighbors because on the billion-scale network, AUC score mainly depends on whether two nodes have neighbors instead of the weights.

\subsubsection{Dynamic Link Prediction}
To simulate the evolving scenarios of real networks, we first randomly split the network into 30\% training and 70\% testing, train embedding vectors on the training set and evaluate the link prediction results on the testing set. Then, we randomly select 10\% (with respect to the whole network) from the testing set as the evolving part and add it to the training set. After updating the embedding vectors using the new training data, we evaluate the link prediction results on the rest of the network. The process is repeated until when 70\% of the network becomes training and the rest 30\% is testing. We adopt two versions of our method: one by dynamic updating as the network evolves (RandNE-D), and one by re-running the algorithm at each time (RandNE-R). To fairly compare the effectiveness of our dynamic updating method and re-running the algorithm, we use the same random seed for RandNE-D and RandNE-R.

The results in Table \ref{table:auclarge2} show that our proposed method again consistently outperforms the baselines on the AUC scores. Besides, RandNE-D shows identical results as RandNE-R, verifying that our dynamic updating has no error aggregation. All methods have larger AUC scores as training data increases because more information is provided.

    \begin{table}[t]
    \caption{AUC scores of dynamic link prediction on WeChat.}\label{table:auclarge2}
    \centering
    \begin{tabular}{ c | c | c | c | c | c}
    \hline
    Observed Edges & 30\% & 40\% & 50\% & 60\% & 70\% \\ \hline
    RandNE-D        & $\mathbf{0.646}$ & $\mathbf{0.689}$ & $\mathbf{0.726}$ & $\mathbf{0.756}$ & $\mathbf{0.780}$\\ \hline
    RandNE-R        & $\mathbf{0.646}$ & $\mathbf{0.689}$ & $\mathbf{0.726}$ & $\mathbf{0.756}$ & $\mathbf{0.780}$\\ \hline
    Common Neighbors & 0.575 & 0.611 & 0.647 & 0.681 & 0.712 \\ \hline
    Adamic Adar     & 0.575 & 0.611 & 0.647 & 0.681 & 0.712 \\ \hline
    Random          & 0.500 & 0.500 & 0.500 & 0.500 & 0.500 \\ \hline
    \end{tabular}
    \end{table}

\subsubsection{Speedup via Distributed Computing}
\begin{figure}[t]
\centering
\includegraphics[width=4.5cm]{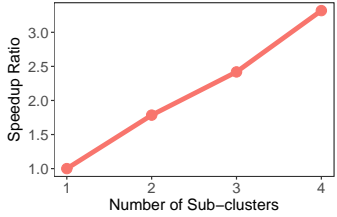}
\caption{The speedup ratio of RandNE via distributed computing.}
\label{fig:ParaSpeedup}
\end{figure}

We evaluate the performance of our method in distributed computing by reporting the speedup ratio. We divide the 16 servers into 4 sub-clusters, with each sub-cluster containing 4 servers. We vary the number of sub-clusters used for distributed computing and record the running time and speedup ratio. Figure \ref{fig:ParaSpeedup} shows that our method has a linear speedup ratio with a slope of approximately 0.8. The slope is slightly less than 1 because of subtle differences in the servers and some extra costs, e.g. reading data. We also report the exact running time in Table \ref{table:time}. It shows that RandNE can learn all the node embeddings of WeChat within 7 hours with 16 normal servers, which is promising for real billion-scale networks.

 \begin{table}[t]
    \caption{The running time of RandNE via distributed computing.}\label{table:time}
    \centering
    \begin{tabular}{ c | c | c | c | c}
    \hline
    Number of Sub-clusters  & 1 & 2 & 3 & 4  \\ \hline
    Running Time(s)     & 82157 & 46029 & 33965 & 24757 \\ \hline
    \end{tabular}
    \end{table}

We also analyze the impact of the proximity order $q$ on WeChat network, which shows similar patterns as moderate-scale networks in Figure \ref{fig:ParaOrder}. We omit the results here for brevity.

\section{Conclusion}
In this paper, we study the problem of embedding billion-scale networks while preserving high-order proximities. We propose RandNE, a novel and simple network embedding method based on random projection and design an iterative projection procedure to efficiently preserve the high-order proximities. Theoretical analysis shows that i) RandNE is more computationally efficient than the existing methods by orders of magnitude, ii) it can well support distributed computing without any communication cost in the calculation, and iii) it can efficiently incorporate the dynamic changes of the networks without error aggregation. Extensive experimental results on multiple datasets with different scales demonstrate the efficiency and efficacy of our proposed method.

One future direction is to generalize this idea to incorporate node attributes and node labels. It is also interesting to explore other random projection methods beyond Gaussian projection.

\section*{Acknowledgment}
This work was supported in part by National Program on Key Basic
Research Project (No. 2015CB352300), National Natural Science
Foundation of China (No. 61772304, No. 61521002, No. 61531006,
No. 61702296), National Natural Science Foundation of China Major
Project (No. U1611461), the research fund of Tsinghua-Tencent Joint
Laboratory for Internet Innovation Technology, and the Young Elite
Scientist Sponsorship Program by CAST. All opinions, findings and conclusions in this paper are those of the
authors and do not necessarily reflect the views of the funding
agencies.


\bibliographystyle{IEEEtran}
\bibliography{bare_conf}
%




\section*{Appendix}
\subsection{Proof of Theorem 1}
\begin{IEEEproof}
From the Johnson-Lindenstrauss property \cite{arriaga2006algorithmic}, for any $\mathbf{s}\in \mathbb{R}^N$, we have
\begin{small}
\begin{equation}\label{eq:JLT}
    P \left[ \left( 1 - \epsilon \right) \left\| \mathbf{s} \right\|^2_2 \leq \left\| \mathbf{s} \cdot \mathbf{R} \right\|^2_2 \leq \left( 1 + \epsilon \right) \left\| \mathbf{s} \right\|^2_2 \right] \geq 1 - 2e^{- \frac{\left( \epsilon^2 - \epsilon^3 \right) d}{4}}.
\end{equation}
\end{small}
Then, we can get:
\begin{equation}
\begin{aligned}
   & P \left[ \; \left\| \mathbf{S} \cdot \mathbf{S}^T - \mathbf{U} \cdot \mathbf{U}^T \right\|_2 > \epsilon \left\| \mathbf{S}^T \cdot \mathbf{S} \right\|_2 \right] \\
    = \; & P \left[ \sup_{\left\| \mathbf{x} \right\|_2 = 1} \left| \left\| \mathbf{x} \cdot \mathbf{S}\right\|_2^2 - \left\| \mathbf{x} \cdot \mathbf{S} \cdot \mathbf{R} \right\|_2^2 \right| > \epsilon \sup_{\left\| \mathbf{x} \right\|_2 = 1}  \left\| \mathbf{x} \cdot \mathbf{S}\right\|_2^2 \right] \\
  \leq \; & P \left[ \exists \; \mathbf{s} \in rowspan \left\{\mathbf{S} \right\}, \left| \left\| \mathbf{s}\right\|_2^2 - \left\| \mathbf{s} \cdot \mathbf{R} \right\|_2^2 \right| > \epsilon \left\| \mathbf{s}\right\|_2^2 \right] \\
  \leq \; & r_{\mathbf{S}} P\left[ \mathbf{s}, \left| \left\| \mathbf{s}\right\|_2^2 - \left\| \mathbf{s} \cdot \mathbf{R} \right\|_2^2 \right| > \epsilon \left\| \mathbf{s}\right\|_2^2 \right] = 2 r_{\mathbf{S}} e^{-\frac{\left( \epsilon^2 - \epsilon^3 \right) d}{4}}.
 \end{aligned}
\end{equation}
The last line is resulted from the union bound and Eq. \eqref{eq:JLT}.
\end{IEEEproof}
\subsection{Proof of Theorem 3}
\begin{IEEEproof}
Denote the number of changed nodes and number of changed edges as $N'$ and $M'$ respectively.

Then, the time complexity from line \ref{step11} to line \ref{step12} is $O(N' \cdot d^2 + N' \cdot q \cdot d)$, i.e. linear with respect to $N'$.

For line \ref{step13} to line \ref{step14}, consider any edge $(s,t)$ in the changes of edges, i.e. $\Delta \mathbf{A}(s,t) \neq 0$. It is easy to see that only the embedding vectors of nodes in its q-step neighborhood, i.e. nodes that are connected to node $s$ or node $t$ with a path of length that is no longer than $q$, will change. Formally, denote the expected size of q-step neighborhood of a randomly chosen edge as $\mathcal{R}(q)$. Then, the expected time complexity from line \ref{step13} to line \ref{step14} is $O(M' \cdot \mathcal{R}(q))$, i.e. linear with respect to $M'$. As for $\mathcal{R}(q)$, it is a constant that depends on the individual network structure. Generalized from \cite{zhang2018timers}, the exact results for two typical types of networks are as follows:
\begin{itemize}
\item For random networks, (i.e. the Erdos Renyi model\cite{erdos1960evolution}), $\mathcal{R}(q) \approx 2 \left(d_{avg} \right) ^q$, where $d_{avg}$ is the average degree of the network.
\item For Barabasi Albert model \cite{barabasi1999emergence}, a widely studied example of preferential attachment networks,  $\mathcal{R}(q) \approx 2 \left[ \frac{6}{\pi^2} \left(\log d_{max} + \gamma \right) \right]^q$, where $d_{max}$ is the maximum degree of the network and $\gamma \approx 0.58$ is the Euler-Mascheroni constant.
\end{itemize}
As a result, the overall time complexity is $O(N' \cdot d^2 + N' \cdot q \cdot d + M' \cdot \mathcal{R}(q))$, which concludes the proof.
\end{IEEEproof}
\end{document}